\documentclass[12pt]{article}
\begin{document}
\title{The Planck Oscillator Model}
\author{B.G. Sidharth\\
International Institute for Applicable Mathematics \& Information Sciences\\
Hyderabad (India) \& Udine (Italy)\\
B.M. Birla Science Centre, Adarsh Nagar, Hyderabad - 500 063
(India)}
\date{}
\maketitle
\begin{abstract}
We consider a model for an underpinning of the universe: there are
oscillators at the Planck scale in the background dark energy.
Starting from a coherent array of such oscillators it is possible to
get a description from elementary particles to Black Holes including
the usual Hawking-Beckenstein theory. There is also a description of
Gravitation in the above model which points to a unified description
with electromagnetism.
\end{abstract}
\section{Introduction}
Max Planck, more than a century ago introduced a combination of the
well known fundamental constants, $\hbar,G,c$ that gave a length,
mass and time scale viz.,
$$l = \sqrt{\frac{\hbar G}{c^3}} \sim 10^{-33}cm$$
\begin{equation}
m = \sqrt{\frac{\hbar c}{G}} \sim 10^{-5}gm\label{e1}
\end{equation}
$$t = \sqrt{\frac{\hbar G}{c^5}} \sim 10^{-42}sec$$
We can easily verify that $l$ plays the role of the Compton length
and the Schwarzchild radius of a black hole of the mass $m$
\cite{kiefer}
\begin{equation}
l = \frac{\hbar}{2mc}, \, l = \frac{2Gm}{c^2}\label{e2}
\end{equation}
Today in various Quantum gravity approaches the Planck length $l$ is
considered to be the fundamental minimum length, and so also the
time interval $t$. Spacetime intervals smaller than given in
(\ref{e1}) and (\ref{e2}) are meaningless both classically and
Quantum mechanically. Classically because we cannot penetrate the
Schwarzchild radius, and Quantum mechanically because we encounter
unphysical phenomena inside a typical Compton scale. All this has
been discussed in greater detail by the author (Cf.ref.\cite{tduniv}
and several references therein). In any case, it is worth pointing
out that Quantum mechanically it is meaningless to speak about
spacetime points, as these would imply infinite momenta and energy.
This is at the root of the infinities and divergences that we
encounter, both in the classical theory of the electron as also in
Quantum mechanics and Quantum Field Theory. In Quantum Field Theory
we have to take recourse to the mathematical device of
Renormalization
to overcome this difficulty.\\
At another level, it may be mentioned that the author's 1997 model
invoked a background dark energy and fluctuations therein to deduce
a model of the universe that was accelerating with a small
cosmological constant, together with several other relations
completely consistent with Astrophysics and Cosmology
(Cf.ref.\cite{bgscu} and several references therein). At that time
it may be recalled, the accepted standard big bang model told us
that the universe was dominated by dark matter and was consequently
decelerating and would eventually come to a halt. However the
observations of distant supernovae by Perlmutter and others
confirmed in 1998 the dark energy driven accelerating universe of
the author. All this is well known.
\section{The Planck Oscillator Model}
It is against this backdrop that the author had put forward his
model of Planck oscillators in the dark energy driven Quantum
vacuum, several years ago (Cf.ref.\cite{uof} and several references
therein, \cite{fpl2000}). To elaborate let us consider an array of
$N$ particles, spaced a distance $\Delta x$ apart, which behave like
oscillators that are connected by springs. As is known we then have
\cite{uof,bgsfpl15,good,vandam} (Cf.in particular ref.\cite{vandam})
$$r = \sqrt{N \Delta x^2}$$
\begin{equation}
ka^2 \equiv k \Delta x^2 = \frac{1}{2} k_B T\label{e3}
\end{equation}
where $k_B$ is the Boltzmann constant, $T$ the temperature, $r$ the
total extension and $k$ is the spring constant given by
\begin{equation}
\omega_0^2 = \frac{k}{m}\label{e4}
\end{equation}
\begin{equation}
\omega = \left(\frac{k}{m} a^2\right)^{\frac{1}{2}} \frac{1}{2} =
\omega_0 \frac{a}{r}\label{e5}
\end{equation}
It must be pointed out that equations (\ref{e3}) to (\ref{e5}) are
general and a part of the well known theory referred to in
\cite{bgsfpl15,good,vandam}. In particular there is no restriction
on the temperature $T$. $m$ and $\omega$ are the mass of the
particle and frequency of oscillation. In (\ref{e4}) $\omega_0$ is
the frequency of the individual oscillator, while in (\ref{e5})
$\omega$ is the frequency of the array of $N$ oscillators, $N$
given in (\ref{e3}).\\
We now take the mass of the particles to be the Planck mass and set
$\Delta x \equiv a = l$, the Planck length as the mass and length
are free parameters. We also use the well known Einstein-de Broglie
relations that give quite generally the frequency in terms of energy
and mass.
\begin{equation}
E = \hbar \omega = mc^2\label{e6}
\end{equation}
It may be immediately observed that if we use (\ref{e4}) and
(\ref{e3}) we can deduce that
$$k_B T \sim mc^2$$
Independently of the above steps this agrees with the (Beckenstein)
temperature of a Black Hole of Planck mass in the usual theory.
Indeed as noted, Rosen \cite{rosen} had shown that a Planck mass
particle at the Planck scale can be considered to be a Universe in
itself with a
Schwarzchild radius equalling the Planck length.\\
Thus we have shown from the above theory of oscillators that an
oscillator with the Planck mass and with a spatial extent at the
Planck scale has the same temperature as the Beckenstein temperature
of a Schwarzchild Black Hole of mass given by the Planck mass. We
may reiterate that while equations (\ref{e3}) to (\ref{e5}) are
valid generally, in the special case where the mass is taken to be
the Planck mass and the distance $a$ is taken to be the Planck
length, we get a complete identification with the corresponding
Schwarzchild Black Hole and the Beckenstein temperature. The above
results can be obtained by a different route as described in
\cite{bgsijmpa}.
\section{Elementary Particles and Black Hole Thermodynamics}
We have also argued that, given the well known effect that the
universe consists of $N \sim 10^{80}$ elementary particles like the
pion, it is possible to deduce that a typical elementary particle
consists of $n \sim 10^{40}$ Planck oscillators. These form a
coherent array described by equations (\ref{e3}) to (\ref{e6})
above. In this case $N$ in (\ref{e3}) becomes $n$ and we can
immediately deduce the following
\begin{equation}
l_\pi = \sqrt{n}l \, m_\pi = \frac{m}{\sqrt{n}}\label{A}
\end{equation}
which give the Compton wavelength and mass of a typical elementary
particle. So a typical elementary particle is given as the lowest
energy state of the above coherent array of $n$ Planck
oscillators.\\
Interestingly the above description can lead to an immediate
correspondence with black hole thermodynamics. We now rewrite
equation (\ref{e5}) as, (interchanging the roles of $\omega$ and
$\omega_0$),
$$\omega_0 = \frac{r}{a} \omega$$
Remembering that, quite generally, the frequency and mass are
related by (\ref{e6}), i.e.,
$$\omega = \frac{mc^2}{\hbar},$$
we get on using (\ref{e3})
\begin{equation}
\hbar \omega \langle \frac{l}{r}\rangle^{-1} \approx mc^2 \times
\frac{r}{l} \approx Mc^2 = \sqrt{N} mc^2\label{e7}
\end{equation}
where we now consider not the lowest energy states of the array as
previously but rather energy states much higher than the Planck
energy. Generally, if an arbitrary mass $M$, as in (\ref{e7}), is
given in terms of $N$ Planck oscillators, in the above model, then
we have from (\ref{e7}) and (\ref{e3}):
\begin{equation}
M = \sqrt{N} m \, \mbox{and} \, R = \sqrt{N} l,\label{e8}
\end{equation}
where $R$ is the radius or extension of the object. We must stress
the factor $\sqrt{N}$ in (\ref{e8}), arising from the fact that the
oscillators are coupled, as given in (\ref{e3}). If the oscillators
had not been coupled, or equivalently had not formed a coherent
system, then we would have, for example, $M = Nm$ or $R = Nl$
instead of (\ref{e8}). Using the fact that $l$ has been chosen to be
the Schwarzchild radius of the mass $m$, this gives immediately,
$$R = 2GM/c^2$$
This shows that if an arbitrary mass $M$ consists of $N$ coherent
Planck oscillators as above, and specifically equation (\ref{e8}),
then its radius $R$ is given by the above expression, which is its
Schwarzchild radius. In other words, such an object shows up as a
Schwarzchild Black Hole. It must be emphasized that the expression
for $R$ follows from the theory of oscillators, specifically
equation (\ref{e8}) and shows that it is identical to the
Schwarzchild radius for the same mass $M$. We have merely used the
known equivalence of the Planck length and Schwarzchild radius for
the Planck mass.
\section{Thermodynamic Gravitation}
We can push the above consideration further. So far we have
considered only a coherent array. This is necessary for meaningful
physics and leads to the elementary particle masses and their other
parameters as seen above. Cercignani \cite{cer} had used Quantum
oscillations, though just before the dark energy era -- these were
the usual Zero Point oscillations, which had also been invoked by
the author in his model. Invoking gravitation, what he proved was,
in his own words, ''Because of the equivalence of mass and energy,
we can estimate that this (i.e. chaotic oscillations) will occur
when the former will be of the order of $G[(\hbar \omega )c^{-2}]^2
[\omega^{-1}c]^{-1} = G\hbar^2\omega^3 c^{-5}$, where $G$ is the
constant of gravitational attraction and we have used as distance
the wavelength. This must be less than the typical electromagnetic
energy $\hbar \omega$. Hence $\omega$ must be less than
$(G\hbar)^{-1/2}c^{5/2}$, which gives a gravitational cut off for
the frequency in the zero-point energy."\\
In other words he deduced that there has to be a maximum frequency
of oscillators given by
\begin{equation}
G\hbar \omega^2_{max} = c^5\label{e10}
\end{equation}
for the very existence of coherent oscillations. We would like to
point out that if we use the well known equation encountered above
namely
$$\hbar \omega = mc^2,$$
in equation (\ref{e10}) we get the well known relation
\begin{equation}
Gm^2_P \approx \hbar c\label{B}
\end{equation}
which shows that at the Planck scale the gravitational and
electromagnetic strengths are of the same order. This is not
surprising because it was the very basis of Cercignani's derivation
-- if indeed the gravitational energy is greater than that given in
(\ref{B}) that is greater than the electromagnetic energy, then the
Zero Point oscillators, which we have called the Planck oscillators
would become chaotic and incoherent -- there would be no physics.\\
Let us now speak only in terms of the background dark energy. We
also use the fact that there is a fundamental minimum spacetime
interval, namely at the Planck scale. Then we can argue that
(\ref{B}) is the necessary and sufficient condition for coherent
Planck oscillators to exist, in order that there be elementary
particles as given by (\ref{A}) and the rest of the requirements for
the meaningful physical universe. In other words gravitational
energy represented by the gravitation constant $G$ given in
(\ref{B}) is a measure of the energy from the Quantum background
that allows a physically meaningful universe -- in this sense it is
not a separate
fundamental interaction.\\
It is interesting that (\ref{B}) also arises in Sakharov's treatment
of gravitation where it is a residual type of an energy
\cite{sakharov,tduniv}.\\
To proceed if we use (\ref{A}) in (\ref{B}) we can easily deduce
\begin{equation}
Gm^2 \approx \frac{e^2}{n} = \frac{e^2}{\sqrt{N}}\label{C}
\end{equation}
where now $N \sim 10^{80}$, the number of particles in the
universe.\\
Equation (\ref{C}) has been known for a long time emperically,
without any fundamental explanation. Here we have deduced it on the
basis of the Planck oscillator model. Equation (\ref{C}) too brings
out the relation between gravitation and the background Zero Point
Field or Quantum vacuum or dark energy. It shows that the
gravitational energy has the same origin as the electromagnetic
energy but is in a sense a smeared out effect over the $N$ particles
of the universe. In the context of the above considerations we can
now even claim that (\ref{C}) gives the desired unified description
of electromagnetism and gravitation.
\section{Black Holes Again}
If we use (\ref{e4}), (\ref{e3}) and (\ref{e10}) we get \cite{ijtp}
$$k_B T = m \omega^2_{max} l^2 = \frac{c^5/G \hbar}{ml^2} = \frac{\hbar
c^3}{Gm},$$ remembering that $l$ by (\ref{e2}) is the Compton
wavelength. That is we get
\begin{equation}
k_B T = \frac{\hbar c^3}{Gm}\label{e11}
\end{equation}
Equation (\ref{e11}) is the well known Beckenstein temperature
formula valid for a Black Hole of arbitrary mass but derived here
for the Planck mass.\\
Can we now generalize equation (\ref{e11}) to the case of a Black
Hole of arbitrary mass, as in the original Beckenstein formula but
using only the characterization of the Black Hole in terms of Planck
oscillators, as above? This is what we will do. In fact to a Black
Hole of mass $M$ characterized in terms of $N$ oscillators as in
equation (\ref{e8}), we associate a Black Hole temperature defined
by
$$\bar{T} = \frac{T}{\sqrt{N}},$$
where $T$ is given in (\ref{e11}). ($N$ here is not necessarily the
number of particles in the universe). Using this with (\ref{e8}) in
(\ref{e11}) we immediately get
\begin{equation}
k_B \bar{T} = \frac{\hbar c^3}{GM}\label{e12}
\end{equation}
Equation (\ref{e12}) which is the analogue of (\ref{e11}) is the
required result. After this identification, we next use the
following known relations for a Schwarzchild Black Hole
\cite{ruffinizang}:
\begin{equation}
dM = TdS, S = \frac{kc}{4\hbar G}A,\label{e13}
\end{equation}
where $T$ is the Black Hole temperature, now identified with
(\ref{e12}), $S$ the entropy and $A$ is the area of the Black Hole.
The area is given by, using (\ref{e8})
\begin{equation}
A = N l^2\label{e14}
\end{equation}
because, this area is $\sim R^2$. Alternatively this shows that
there are $N$ elementary areas $l^2$ forming the Black Hole. Indeed
this defines the basic quantum of area of quantum gravity approaches
and is in pleasing agreement with the result of Baez deduced from a
different quantum gravity consideration \cite{baez}.\\
Using equations (\ref{e8}), (\ref{e11}) and (\ref{e14}), we can
easily see that equation (\ref{e13}) is valid for the mass $M$ given
by (\ref{e8}) or (\ref{e7}).\\
This completes the identification of Black Holes characterized by
coherent Planck oscillators, with the conventional
Hawking-Beckenstein theory.
\section{Conclusion}
We have shown that it is possible to consider the universe to have
an underpinning of oscillators in the background dark energy. This
leads to a meaningful description of the universe of elementary
particles and also of  black hole thermodynamics. Finally it
provides a description of gravitation, not as a separate fundamental
interaction, but rather as the energy of the background Quantum
vacuum that is a result of the fact that there is a minimum
fundamental spacetime interval in the universe.

\end{document}